\documentclass[twocolumn,nofootinbib,floatfix,amsmath,amssymb,showpacs,superscriptaddress]{revtex4}

\usepackage{graphicx}
\usepackage{dcolumn}
\usepackage{bm}
\usepackage{color}
\def\xslash{x\!\!\!\slash }
\def\qslash{q\!\!\!\slash }

\begin{document}

\title{Isovector axial-vector form factors of octet baryons in QCD}

\author{G\"{u}ray Erkol}
\affiliation{Laboratory for Fundamental Research, Ozyegin University, Kusbakisi Caddesi No:2 Altunizade, Uskudar Istanbul 34662, Turkey}
\email{guray.erkol@ozyegin.edu.tr}
\author{Altug Ozpineci}%
\affiliation{Physics Department, Middle East Technical University, 06531 Ankara, Turkey}
\email{ozpineci@metu.edu.tr}
\date{\today}

\begin{abstract}
We compute the diagonal isovector axial-vector as well as induced pseudoscalar form factors of nucleon, $\Sigma$ and $\Xi$ baryons by employing the light-cone QCD sum rules to leading order in QCD and including distribution amplitudes up to twist 6. Extrapolating our sum-rules results to low-momentum transfers, we make a comparison with experimental and lattice-QCD results where we can achieve a nice qualitative and quantitative agreement.

\end{abstract}
\maketitle

\section{Introduction}
Form factors are important in hadron physics as they provide information about the structure, in particular the shape and the size, of the hadron. The baryon matrix elements of the axial-vector current are parameterized in terms of the axial ($G_{A,B}$) and the induced pseudoscalar ($G_{P,B}$) form factors as follows:
\begin{align}
	\begin{split}
	&\langle B(p^\prime)|A_\mu|B(p)\rangle\\
	&=\bar{u}_B(p^\prime)\left[\gamma_\mu \gamma_5 G_{A,B}(q^2)+\frac{q^\mu}{2m_B}\gamma_5 G_{P,B}(q^2) \right] u_B(p),
	\end{split}
\end{align}
where $A_\mu=\frac12 (\bar{u}\gamma_\mu\gamma_5 u - \bar d \gamma_\mu \gamma_5 d) $ is the isovector axial-vector current, $q=p^\prime-p$ is the momentum transfer and $m_B$ is the baryon mass. Among all, the nucleon form factors have received much attention. The nucleon axial charge, which corresponds to the value of the form factor at zero-momentum transfer ($Q^2=-q^2=0$), can be precisely determined from nuclear $\beta$-decay (the modern value is $g_{A,N}=1.2694(28)$~\cite{Nakamura:2010zzi}). The $Q^2$ dependence of the axial-vector form factor of the nucleon has been studied up to 1~GeV$^2$ from antineutrino scattering~\cite{Ahrens:1988rr} and for $Q^2<0.2$~GeV$^2$ from pion electro-production on the proton~\cite{Choi:1993vt}. In the high-$Q^2$ region ($Q^2>2$~GeV$^2$), we have very small amount of relatively old data~\cite{Baker:1981su}. Our information about hyperon axial-vector form factors from experiment is also limited. However, both the low-$Q^2$ ($Q^2<2$~GeV$^2$) and the high-$Q^2$ ($Q^2>2$~GeV$^2$) regions will be accessible by higher-energy experiments such as Miner$\nu$a at Fermilab, which will give a complete understanding of form factors in a wide range of $Q^2$~\cite{Drakoulakos:2004gn}. In these experiments, strangeness-production processes will be able to probe the hyperon form factors with precision. On the theoretical side, there exist some estimates for the axial charges of the hyperons from chiral perturbation theory ($\chi$PT)~\cite{Savage:1996zd, Jiang:2008aqa, Jiang:2009sf}, large N$_c$ limit~\cite{FloresMendieta:1998ii} of QCD and QCD sum rules~(QCDSR)~\cite{Chiu:1985ka}. 

As for the induced pseudoscalar form factor, a recent result from muon-capture experiment predicts $G_{P,N}(q^2=-0.88~m_\mu^2)=7.3 \pm 1.1$~\cite{Andreev:2007wg}, where $m_\mu$ is the muon mass. There exist theoretical results from heavy-baryon $\chi$PT as $g_{P,N}=8.26 \pm 0.16$~\cite{Bernard:2001rs} in consistency with experiment. The prediction from manifestly-invariant $\chi$PT is $g_{P,N}=8.29^{+0.24}_{-0.13}\pm 0.50$~\cite{Schindler:2006it}, where the first and the second errors are due to empirical quantities and truncation in the chiral expansion, respectively.

Concurrently, the lattice calculations provide a first-principles description of hadronic phenomena, which also serve as a valuable tool to determine the hadron couplings and form factors in a model-independent way. While systematic errors such as the finite lattice size and relatively heavy quark masses still exist, the developing technology of the lattice method shows promising advances in removing sources of these errors. Lattice QCD calculations of the axial charge and form factors of the nucleon have reached a mature level~\cite{Edwards:2005ym, Khan:2006de, Hagler:2007xi, Alexandrou:2007zz, Yamazaki:2008py}. While it is difficult to measure hyperon properties experimentally due to their short lifetimes, the method of lattice QCD makes it possible to extract such information. Namely, there have been recent attempts to extract the hyperon axial charges and meson couplings using lattice QCD~\cite{Lin:2007ap, Sasaki:2008ha, Erkol:2008yj, Erkol201036}. Simulations with more realistic setups with smaller lattice spacing and larger lattice size employing much lighter quarks and a dynamical $s$-quark are under way, which will also provide valuable information about hyperon form factors at high momentum transfers.

A complementary approach to lattice QCD is the method of QCD sum rules, which is a powerful tool to extract qualitative and quantitative information about hadron properties~\cite{Shifman:1978bx, Shifman:1978by, Reinders:1984sr, Ioffe:1983ju}. In this approach, one starts with a correlation function that is constructed in terms of the interpolating fields, which are chosen with respect to the quantum numbers of the hadron in question. In the traditional method one proceeds with the calculation of the correlation function using the Operator Product Expansion (OPE), which is formulated with Wilson coefficients and local operators in terms of the nonperturbative structure of the QCD vacuum, in the deep Euclidian region. This correlation function is matched with an {\em Ansatz} that is introduced in terms of hadronic degrees of freedom on the phenomenological side. The matching provides a determination of hadronic parameters like baryon masses, magnetic moments, coupling constants of hadrons, and so on.

One alternative to the traditional method as far as the hadron interactions at moderately large momentum transfers are concerned is the light-cone sum rules (LCSR)~\cite{Braun:1988qv, Balitsky:1989ry, Chernyak:1990ag}. In this technique, the light-cone kinematics at $x^2\rightarrow 0$ governs the asymptotic behavior of the correlation function. The singularity of the Wilson coefficients is determined by the twist of the corresponding operator. Then using the moments of the baryon distribution amplitudes (DAs), one can calculate the relevant hadron matrix elements.

LCSR have proved to be rather successful in extracting the values of the hadron form factors at high-momentum transfers. In Ref.~\cite{Braun:2006hz}, the electromagnetic and the axial form factors of the nucleon have been calculated to leading order and with higher-twist corrections. It has been found that a light-cone formulation of the nucleon DAs gives a description of the experimental data rather well. This calculation has been generalized to isoscalar and induced pseudoscalar axial-vector form factors of the nucleon in Refs.~\cite{Aliev:2007qu, Wang:2006su}.

Our information about the DAs of the octet hyperons were scarce and as a result not much effort has been spent on these baryons. However, the DAs of octet hyperons have recently become available and their electromagnetic form factors have been calculated by Liu \emph{et al.}~\cite{Liu:2008yg, Liu:2009uc}. Motivated by these advances in formulating the SU(3) sector in LCSR and ongoing simulations in lattice QCD to give a first-principles description of hadron interactions, in this work we study the axial-vector form factors of strange octet baryons using LCSR. Note that the axial-vector current is anomalous in QCD. Although this anomaly cancels in the isovector channel, it might have a significant contribution in the isoscalar channel. Since a study of the isoscalar axial-vector form factor would be unreliable without the inclusion of the anomaly effects, in this work we restrict our attention to the isovector form factors. To this end, we compute the diagonal isovector as well as the induced pseudoscalar form factors of nucleon, $\Sigma$ and $\Xi$ baryons by employing their recently extracted DAs. Our paper is organized as follows: In the following section, we give the formulation of the baryon form factors on the light cone and derive our sum rules. In Section 3, we present our numerical results and in the last section, we conclude our work with a discussion on our results.

\section{Formulation of baryon axial form factors}
In the LCSR method one starts with the following two-point correlation function:
\begin{equation}\label{corrf}
	\Pi^{B}_\mu(p,q)=i\int d^4 x e^{iqx} \langle 0 |T[\eta_B(0)A_\mu(x)]|B(p)\rangle,
\end{equation}
where $\eta_B(x)$ are the baryon interpolating fields for the $N$, $\Sigma$, $\Xi$. There are several local operators with the quantum numbers of spin-1/2 baryons one can choose from. Here we work with the general form of the interpolating fields parameterized as follows for the $N$,$\Sigma$ and $\Xi$:
\begin{align}\label{intf}
	\begin{split}
	\eta_N=&2\epsilon^{abc}\sum_{\ell=1}^{2}(u^{aT}(x) C J_1^\ell d^b(x))J_2^\ell u^c(x),	\\
	\eta_\Sigma=&\eta_N(d\rightarrow s),\\
	\eta_\Xi=&\eta_N(u\rightarrow s,\, d\rightarrow u),\\
	\end{split}
\end{align} 
with $J_1^1=I$, $J_1^2=J_2^1=\gamma_5$ and $J_2^2=\beta$, which is an arbitrary parameter that fixes the mixing of two local operators. We would like to note that when the choice $\beta=-1$ is made the interpolating fields above give what are known as Ioffe currents for baryons. Here $u(x)$, $d(x)$ and $s(x)$ denote the $u$-, $d$- and $s$- quark fields, respectively, $a$, $b$, $c$ are the color indices and $C$ denotes charge conjugation.

The short-distance physics corresponding to high momenta $p^{\prime 2}$ and $q^2$ is calculated in terms of quark and gluon degrees of freedom. Inserting the interpolating fields in Eq.~\eqref{intf} into the correlation function in Eq.~\eqref{corrf}, we obtain
\begin{widetext}
\begin{align}\label{corrfunc}
	\begin{split}
	\Pi^B_\mu=&\frac{1}{2}\int d^4 x e^{iqx}\sum^2_{\ell=1}\left\{c_1(C J_1^\ell)_{\alpha\gamma} \left[J_2^\ell S(-x)\gamma_\mu \gamma_5\right]_{\rho\beta}4\epsilon^{abc}\langle 0|q_{1\alpha}^a(0) q_{2\beta}^b(x) q_{3\gamma}^c(0)|B\rangle\right.\\
	&+ c_2 (J_2^\ell)_{\rho\alpha}\left[(CJ_1^\ell)^T S(-x)\gamma_\mu \gamma_5\right]_{\gamma\beta}4\epsilon^{abc}\langle 0|q_{1\alpha}^a(x) q_{2\beta}^b(0) q_{3\gamma}^c(0)|B\rangle\\
	&\left. + c_3 (J_2^\ell)_{\rho\beta}\left[CJ_1^\ell S(-x)\gamma_\mu \gamma_5\right]_{\alpha\gamma}4\epsilon^{abc}\langle 0|q_{1\alpha}^a(0) q_{2\beta}^b(0) q_{3\gamma}^c(x)|B\rangle\right\},
	\end{split}
\end{align}
\end{widetext}
where $q_{1,2,3}$ denote the quark fields and $c_{1,2,3}$ are constants which will be determined according to the baryon in question. $S(x)$ represents the light-quark propagator
\begin{equation}\label{qprop}
	S(x)=\frac{i\xslash}{2\pi^2x^4}-\frac{\langle q\bar{q}\rangle}{12}\left(1+\frac{m_0^2 x^2}{16}\right).
\end{equation}
Here the first term gives the hard-quark propagator. The second term represents the contributions from the nonperturbative structure of the QCD vacuum, namely, the quark and quark-gluon condensates. These contributions are removed by Borel transformations as will be explained below. We note that the hard-quark propagator receives corrections in the background gluon field, which are expected to give negligible contributions as they are related to four- and five-particle baryon distribution amplitudes~\cite{Diehl:1998kh}. Following the common practice, in this work we shall not take into account such contributions, which leaves us with only the first term in Eq.~\eqref{qprop} to consider.

The matrix elements of the local three-quark operator 
\[4\epsilon^{abc}\langle 0|q_{1\alpha}^a(a_1 x) q_{2\beta}^b(a_2 x) q_{3\gamma}^c(a_3 x)|B\rangle\] ($a_{1,2,3}$ are real numbers denoting the coordinates of the valence quarks) can be expanded in terms of DAs using the Lorentz covariance, the spin and the parity of the baryon. Based on a conformal expansion using the approximate conformal invariance of the QCD Lagrangian up to 1-loop order, the DAs are then decomposed into local nonperturbative parameters, which can be estimated using QCD sum rules or fitted so as to reproduce experimental data. We refer the reader to Refs.~\cite{Braun:2006hz, Liu:2008yg, Liu:2009uc} for a detailed analysis on DAs of $N$, $\Sigma$, $\Xi$, which we employ in our work to extract the axial-vector form factors.

The long-distance side of the correlation function is obtained using the analyticity of the correlation function, which allows us to write the correlation function in terms of a dispersion relation of the form
\[\Pi^B_{\mu}(p,q)=\frac{1}{\pi}\int_0^\infty \frac{\text{Im}\Pi^B_\mu(s)}{(s-p^{\prime 2})}ds\]
The ground-state hadron contribution is singled out by utilizing the zero-width approximation
\[\text{Im}~\Pi^B_\mu=\pi \delta(s-m_B^2)\langle 0|\eta^B|B(p^\prime)\rangle \langle B(p^\prime)|A_\mu|B(p)\rangle + \pi \rho^h(s)\]
and by expressing the correlation function as a sharp resonance plus continuum which starts above the continuum threshold, $s_0$, i.e. $\rho^h(s)=0$ for $s < s_0$. The matrix element of the interpolating current between the vacuum and baryon state is defined as
\[\langle 0|\eta^B|B(p,s)\rangle=\lambda_B\upsilon(p,s)\]
where $\lambda_B$ is the baryon overlap amplitude and $\upsilon(p,s)$ is the baryon spinor.

The QCD sum rules are obtained by matching the short-distance calculation of the correlation function with the long-distance calculation. Using the most general decomposition of the matrix element~(see Eq.~(2.3) in Ref.~\cite{Braun:2000kw}) and taking the Fourier transformations we obtain
\begin{widetext}
	\begin{align}
		\begin{split}
			-&\frac{\lambda_B}{m_B^2-p^{\prime 2}}G_{A,B}\\
			=&\frac{1}{2}\left\{m_B\int^1_0 \frac{dt_2}{(q-p t_2)^2}\left[(1-\beta)F_1(t_2)+(1+\beta) F_2(t_2)\right] + m_B\int^1_0 \frac{dt_3}{(q-p t_3)^2}\left[(1-\beta)F_3(t_3)+(1+\beta) F_4(t_3)\right]\right. \\
			&+m_B^3 \int^1_0 \frac{dt_2}{(q-p t_2)^4}\left[(1-\beta)F_5(t_2)+(1+\beta) F_6(t_2)\right] + m_B^3 \int^1_0 \frac{dt_3}{(q-p t_3)^4}\left[(1-\beta)F_7(t_3)+(1+\beta) F_8(t_3)\right]\\
			&\left. +m_B^3 \int^1_0 \frac{dt_2}{(q-p t_2)^4}\left[(1-\beta)F_9(t_2)+(1+\beta) F_{10}(t_2)\right] + m_B^3 \int^1_0 \frac{dt_3}{(q-p t_3)^4}\left[(1-\beta)F_{11}(t_3)+(1+\beta) F_{12}(t_3)\right]\right\}
		\end{split}
	\end{align}
for the axial-vector form factors at structure $\qslash\gamma_\mu\gamma_5$ and
	\begin{align}
		\begin{split}
			-&\frac{\lambda_B}{m_B^2-p^{\prime 2}}G_{P,B}\\
			=&\frac{1}{2}\left\{m_B^2 \int^1_0 \frac{dt_2}{(q-p t_2)^4}\left[(1-\beta)F_{13}(t_2)+(1+\beta) F_{14}(t_2)\right] + m_B^2 \int^1_0 \frac{dt_3}{(q-p t_3)^4}\left[(1-\beta)F_{15}(t_3)+(1+\beta) F_{16}(t_3)\right]\right\}
		\end{split}
	\end{align}
for the induced pseudoscalar form factor at the structure $q^\mu\qslash\gamma_5$. The explicit form of the functions that appear in the above sum rules are given in terms of DAs as follows: 
{\allowdisplaybreaks
	\begin{align*}
			&F_1=\int_0^{1-t_2} dt_1 \left[c_1(-A_2-A_3+V_2-V_3)+c_2(A_1+V_1)\right](t_1,t_2,1-t_1-t_2),\\
			&F_2=\int_0^{1-t_2} dt_1 \left[c_1(P_1+S_1+2T_2+T_3-T_7)+c_2(P_1+S_1+T_3-T_7)\right](t_1,t_2,1-t_1-t_2),\\
			&F_3=\int_0^{1-t_3} dt_1 \left[c_3(A_1-V_1)\right](t_1,1-t_1-t_3,t_3),\\
			&F_4=\int_0^{1-t_3} dt_1 \left[c_3(P_1+S_1-T_3+T_7)\right](t_1,1-t_1-t_3,t_3),\\
			&F_5=\int_0^{1-t_2} dt_1 \left[c_1(V_1^M-A_1^M)+c_2(V_1^M+A_1^M)\right](t_1,t_2,1-t_1-t_2),\\
			&F_6=\int_0^{1-t_2} dt_1 \left[c_1(3 T_1^M)+c_2(T_1^M)\right](t_1,t_2,1-t_1-t_2),\\
			&F_7=\int_0^{1-t_3} dt_1 \left[c_3(A_1^M-V_1^M)\right](t_1,1-t_1-t_3,t_3),\\
			&F_8=\int_0^{1-t_3} dt_1 \left[-c_3 T_1^M\right](t_1,1-t_1-t_3,t_3),\\
			&F_9=\int_1^{t_2}d\lambda \int_1^\lambda d\rho \int_0^{1-\rho} dt_1 \left[(c_1+c_2) (A_1-A_2+A_3+A_4-A_5+A_6)\right.\\[0.5ex]
			&\left.\qquad\qquad\qquad+ (c_2-c_1) (V_1-V_2-V_3-V_4-V_5+V_6)\right](t_1,\rho,1-t_1-\rho),\\[0.7ex]
			&F_{10}=\int_1^{t_2}d\lambda \int_1^\lambda d\rho \int_0^{1-\rho} dt_1 \left[c_1 (-3T_1+T_2+2T_3+T_4+T_5-3T_6+4T_7+4T_8)\right.\\[0.5ex]
			&\left.\qquad\qquad\qquad + c_2 (-T_1-T_2+2T_3+2T_4-T_5-T_6)\right](t_1,\rho,1-t_1-\rho),\\[0.7ex]
			&F_{11}=\int_1^{t_3}d\lambda \int_1^\lambda d\rho \int_0^{1-\rho} dt_1 \left[c_3(A_1-A_2+A_3+A_4-A_5+A_6\right.\\
			&\left.\qquad\qquad\qquad-V_1+V_2+V_3+V_4+V_5-V_6)\right] (t_1,1-t_1-\rho,\rho),\\
			&F_{12}=\int_1^{t_3}d\lambda \int_1^\lambda d\rho \int_0^{1-\rho} dt_1 \left[c_3(T_1+T_2-2T_3-2T_4+T_5+T_6)\right] (t_1,1-t_1-\rho,\rho),\\
			&F_{13}=\int_1^{t_2}d\rho \int_0^{1-\rho} dt_1 \left[c_1(A_2+A_3-A_4-A_5-V_2+V_3-V_4+V_5)\right.\\
			&\left.\qquad\qquad\qquad+c_2(A_1+A_3-A_5+V_1-V_3-V_5) \right](t_1,\rho,1-t_1-\rho),\\
			&F_{14}=\int_1^{t_2}d\rho \int_0^{1-\rho} dt_1 \left[c_1(-P_1+P_2-S_1+S_2-2T_2-T_3+T_4+2T_5+T_7-T_8)\right.\\
			&\left.\qquad\qquad\qquad+2c_2(-T_3+T_5+T_7) \right](t_1,\rho,1-t_1-\rho),\\
			&F_{15}=\int_1^{t_3}d\rho \int_0^{1-\rho} dt_1 \left[c_3(A_1-A_2+A_4-V_1+V_2+V_4) \right](t_1,1-t_1-\rho,\rho),\\
			&F_{16}=\int_1^{t_3}d\rho \int_0^{1-\rho} dt_1 \left[c_3(-P_1+P_2-S_1+S_2-2T_2+T_3+T_4-T_7-T_8) \right](t_1,1-t_1-\rho,\rho),\\
	\end{align*}
}
We make the following replacements in order to obtain the sum rule for each baryon we consider:
\begin{align*}
	&G_{N}:\{c_1=c_2=1,\,c_3=-1,\, q_1\rightarrow u,\, q_2\rightarrow u,\, q_3\rightarrow d\},\\
	&G_{\Sigma}:\{c_1=c_2=1,\,c_3=0,\, q_1\rightarrow u,\, q_2\rightarrow u,\, q_3\rightarrow s\},\\
	&G_{\Xi}:\{c_1=c_2=0,\,c_3=1,\, q_1\rightarrow s,\, q_2\rightarrow s,\, q_3\rightarrow d\},
\end{align*}
Note that, in the final sum rules expression, the quarks do not appear explicitly but only implicitly through the DA's, masses and the residues of the corresponding baryons. Thus these replacements simply instruct to use the DA's, mass and residue of the corresponding baryon. They apply to both axial-vector and induced pseudoscalar form factors.

The Borel transformation is performed to eliminate the subtraction terms in the spectral representation of the correlation function. As a result of Borel transformation,  contributions from excited and continuum states are also exponentially suppressed. The contributions of the higher states and the continuum are modeled using the quark-hadron duality and subtracted. Both of the Borel transformation and the subtraction of the higher states are carried out using the following substitution rules (see e.g. \cite{Braun:2006hz}):
\begin{align}
	\begin{split}
		&\int dx \frac{\rho(x)}{(q-xp)^2}\rightarrow -\int_{x_0}^1\frac{dx}{x}\rho(x) e^{-s(x)/M^2},\\
		&\int dx \frac{\rho(x)}{(q-xp)^4}\rightarrow \frac{1}{M^2} \int_{x_0}^1\frac{dx}{x^2}\rho(x) e^{-s(x)/M^2}+\frac{\rho(x)}{Q^2+x_0^2 m_B^2} e^{-s_0/M^2},
	\end{split}
\end{align}
where
\[s(x)=(1-x)m_B^2+\frac{1-x}{x}Q^2,\]
$M$ is the Borel mass and $x_0$ is the solution of the quadratic equation for $s=s_0$:
\[x_0=\left[\sqrt{(Q^2+s_0-m_B^2)^2+4m_B^2(Q^2)}-(Q^2+s_0-m_B^2)\right]/(2m_B^2),\] where $s_0$ is the continuum threshold.

\begin{figure}[t]
	\includegraphics[width=0.9\textwidth]{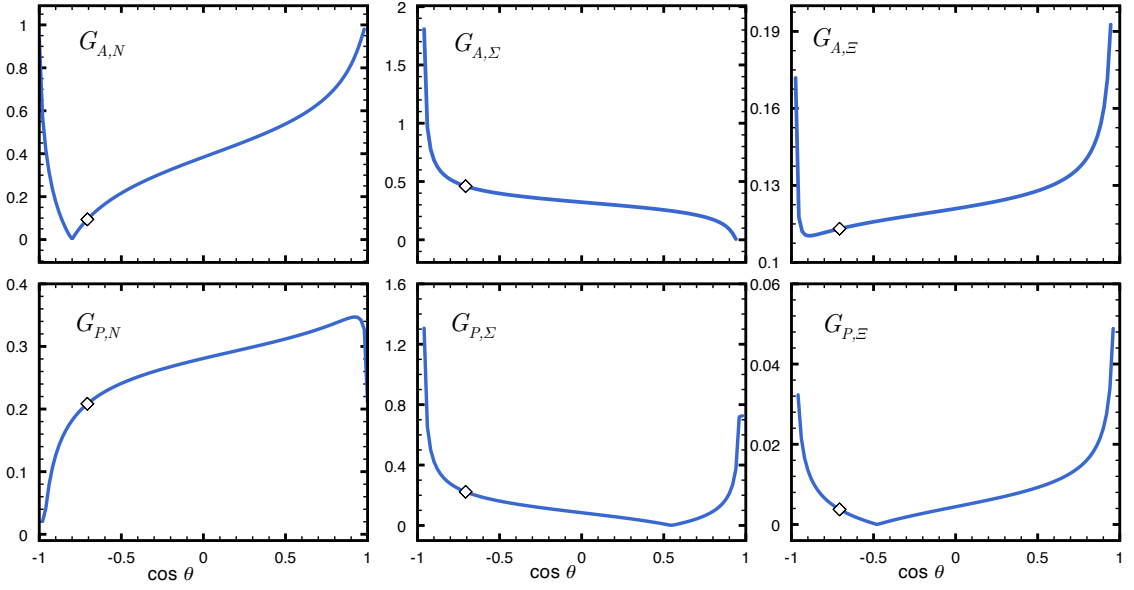}
	\caption{\label{costhq} The form factors as a function of $\cos\theta$. The diamonds mark the points for Ioffe current.}
\end{figure}	


Finally, we obtain the following sum rules for the axial-vector and induced pseudoscalar form factors, respectively: 
	\begin{align}
		\begin{split}
			G_A=&-\frac{1}{2\lambda_B}e^{m_B^2/M^2}\left\{-m_B \int^1_{x_0} \frac{dt_2}{t_2} e^{-s(t_2)/M^2} \left[(1-\beta)F_1(t_2)+(1+\beta) F_2(t_2)\right]\right.\\
			-&m_B \int^1_{x_0} \frac{dt_3}{t_3} e^{-s(t_2)/M^2} \left[(1-\beta)F_3(t_3)+(1+\beta) F_4(t_3)\right]\\
			+&\frac{m_B^3}{M^2} \int^1_{x_0} \frac{dt_2}{t_2^2} e^{-s(t_2)/M^2} \left[(1-\beta)F_5(t_2)+(1+\beta) F_6(t_2)\right] + \frac{m_B^3}{q^2+x_0^2 m_B^2} e^{-s_0/M^2} \left[(1-\beta)F_5(x_0)+(1+\beta) F_6(x_0)\right]\\
			+&\frac{m_B^3}{M^2} \int^1_{x_0} \frac{dt_3}{t_3^2} e^{-s(t_3)/M^2} \left[(1-\beta)F_7(t_3)+(1+\beta) F_8(t_3)\right] + \frac{m_B^3}{q^2+x_0^2 m_B^2} e^{-s_0/M^2} \left[(1-\beta)F_7(x_0)+(1+\beta) F_8(x_0)\right]\\
			+&\frac{m_B^3}{M^2} \int^1_{x_0} \frac{dt_2}{t_2^2} e^{-s(t_2)/M^2} \left[(1-\beta)F_9(t_2)+(1+\beta) F_{10}(t_2)\right] + \frac{m_B^3}{q^2+x_0^2 m_B^2} e^{-s_0/M^2} \left[(1-\beta)F_9(x_0)+(1+\beta) F_{10}(x_0)\right]\\
			+&\left.\frac{m_B^3}{M^2} \int^1_{x_0} \frac{dt_3}{t_3^2} e^{-s(t_3)/M^2} \left[(1-\beta)F_{11}(t_3)+(1+\beta) F_{12}(t_3)\right] + \frac{m_B^3}{q^2+x_0^2 m_B^2} e^{-s_0/M^2} \left[(1-\beta)F_{11}(x_0)+(1+\beta) F_{12}(x_0)\right]
			\right\},
		\end{split}
	\end{align}
	
	\begin{align}
		\begin{split}
			G_P=&-\frac{1}{\lambda_B}e^{m_B^2/M^2}\left\{\frac{m_B^2}{M^2} \int^1_{x_0} \frac{dt_2}{t_2^2} e^{-s(t_2)/M^2}\left[(1-\beta)F_{13}(t_2) +(1+\beta)F_{14}(t_2)\right]\right.\\
			&+ \frac{m_B^2}{Q^2+x_0^2 m_B^2} e^{-s_0/M^2} \left[(1-\beta)F_{13}(x_0)+(1+\beta) F_{14}(x_0)\right]\\
			&+\frac{m_B^2}{M^2} \int^1_{x_0} \frac{dt_3}{t_3^2} e^{-s(t_2)/M^2}\left[(1-\beta)F_{15}(t_3) +(1+\beta)F_{16}(t_3)\right]\\
			 & \left.+ \frac{m_B^2}{Q^2+x_0^2 m_B^2} e^{-s_0/M^2} \left[(1-\beta)F_{15}(x_0)+(1+\beta) F_{16}(x_0)\right]
			\right\}.
		\end{split}
	\end{align}

To obtain a numerical prediction for the form factors, the residues,
$\lambda_B$ are also required. The residues can be obtained from the mass sum rules, and the residue of the $\Sigma$ is given by \cite{Aliev:2010aj}:
\begin{eqnarray}
\label{residue}
\lambda_{\Sigma}^2 e^{-m_{\Sigma^0}^2/M^2} &=&
{M^6\over 1024 \pi^2} (5 + 2 \beta + 5 \beta^2) E_2(x) - {m_0^2\over 96 M^2} (-1+\beta)^2 \langle 
\bar{q} q \rangle^2  \nonumber \\
&-& {m_0^2\over 8 M^2} (-1+\beta^2) \langle  \bar{s} s \rangle  \langle  \bar{q} q \rangle 
+ {3 m_0^2\over 64 \pi^2} (1-\beta^2) \ln \frac{M^2}{\Lambda^2}\left[ m_s \langle  \bar{q} q \rangle + m_q \langle  \bar{s} s \rangle  \right] \nonumber \\
&+&  {3\over 64 \pi^2} (1+\beta)^2 M^2 m_q \langle  \bar{q} q \rangle E_0(x) 
- {3 M^2\over 32 \pi^2} (-1+\beta^2) \left[ m_s \langle \bar{q} q \rangle + m_q \langle \bar{s} s \rangle \right] E_0(x) \nonumber \\
&+& {M^2\over 128 \pi^2}  (5 + 2 \beta + 5 \beta^2) m_s \langle \bar{s} s \rangle E_0(x) 
+ {1\over 24} \left[ 6 (-1+\beta^2) \langle \bar{s} s \rangle \langle \bar{q} q \rangle + (-1+\beta^2) \langle \bar{q} q \rangle^2 \right] \nonumber \\
&+& {m_0^2\over 128 \pi^2} (-1+\beta)^2 m_q \langle \bar{q} q \rangle
+ {m_0^2\over 128 \pi^2} (-1+\beta^2) \left[ 13 m_s \langle \bar{q} q \rangle + 11 m_q \langle \bar{s} s \rangle \right] \nonumber \\
&-& {m_0^2\over 96 \pi^2}(1+\beta+\beta^2) \left( m_q \langle \bar{q} q \rangle -  m_s \langle \bar{s} s \rangle\right)~, 
\end{eqnarray}
\end{widetext}
where $x = s_0/M^2$, and
\begin{eqnarray}
\label{nolabel}
E_n(x)=1-e^{-x}\sum_{i=0}^{n}\frac{x^i}{i!}~. \nonumber
\end{eqnarray}
The residues for the nucleon and $\Xi$ can be obtained from Eq. \eqref{residue}. $\lambda_N^2 e^{-m_N^2/M^2}$ can be obtained by setting $m_s \rightarrow m_q$ and $\langle \bar s s \rangle \rightarrow  \langle \bar q q \rangle$, and $\lambda_\Xi^2e^{-m_\Xi^2/M^2}$ by the exchanges $m_q \leftrightarrow m_s$ and $\langle \bar s s \rangle \leftrightarrow \langle \bar q q \rangle$. We use the following parameter values: $\langle \bar{q}q\rangle=0.8 \langle \bar{s}s\rangle=-(0.243)^3$~GeV$^3$, $m_s=0.14$~GeV, $m_q=0$, $m_0^2=0.8$~GeV$^2$, $\Lambda=0.2$~GeV, $m_N=0.94$~GeV, $m_\Sigma=1.2$~GeV and $m_\Xi=1.3$~GeV. 

\section{Numerical Results and Discussion}
In this section we give our numerical results for the axial-vector form factors of $N$, $\Sigma$ and $\Xi$. For this purpose we need the numerical values of the baryon DAs. The DAs of the nucleon are given in Ref.~\cite{Braun:2006hz} as  expressed in terms of some nonperturbative parameters which are calculated using QCDSR or phenomenological models (see also Ref.~\cite{Lenz:2009ar} for a comparison of nucleon DAs as determined on the lattice~\cite{Gockeler:2008xv} and with other approaches). In this work, we give our results using the parameter set known as Chernyak-Zhitnitsky-like model of the DAs (see Ref.~\cite{Braun:2006hz} for details). As for the DAs of $\Sigma$ and $\Xi$ we use the parameter values as calculated recently by Liu \emph{et al.}~\cite{Liu:2008yg, Liu:2009uc}. In Table~\ref{parameter_table} we list the values of the input parameters entering the DAs of each baryon.

\begin{table}[t]
	\caption{The values of the parameters entering the DAs of $N$, $\Sigma$ and $\Xi$. The upper panel shows the dimensionful parameters for each baryon. In the lower panel we list the values of the five parameters that determine the shape of the DAs, which have been extracted for nucleon only. For $\Sigma$ and $\Xi$ these parameters are taken as zero.}
	\addtolength{\tabcolsep}{10pt}
	\setlength{\extrarowheight}{1.5pt} 
\begin{tabular}{ccccccc}
		\hline\hline 
		Parameter & $N$ & $\Sigma$ & $\Xi$& \\[0.5ex]
		\hline 
		 $f_B$~(GeV$^2$)       & 0.005  & 0.0094 & 0.0099   \\
		 $\lambda_1$~(GeV$^2$) & -0.027 & -0.025 & -0.028   \\
		 $\lambda_2$~(GeV$^2$) & 0.054  & 0.044  & 0.052  \\[0.8ex]
		\end{tabular}	
\begin{tabular}{ccccccc}
			\hline	\hline
		 $V_1^d$ & $A_1^u$ & $f_1^d$ & $f_2^d$ & $f_1^u$  \\[0.5ex]
		\hline 
		 0.23 & 0.38 & 0.40 & 0.22 & 0.07 &  \\[0.5ex]
		\hline\hline
	\end{tabular}
	\label{parameter_table}
\end{table}

\begin{figure}[th]
	\includegraphics[width=0.5\textwidth]{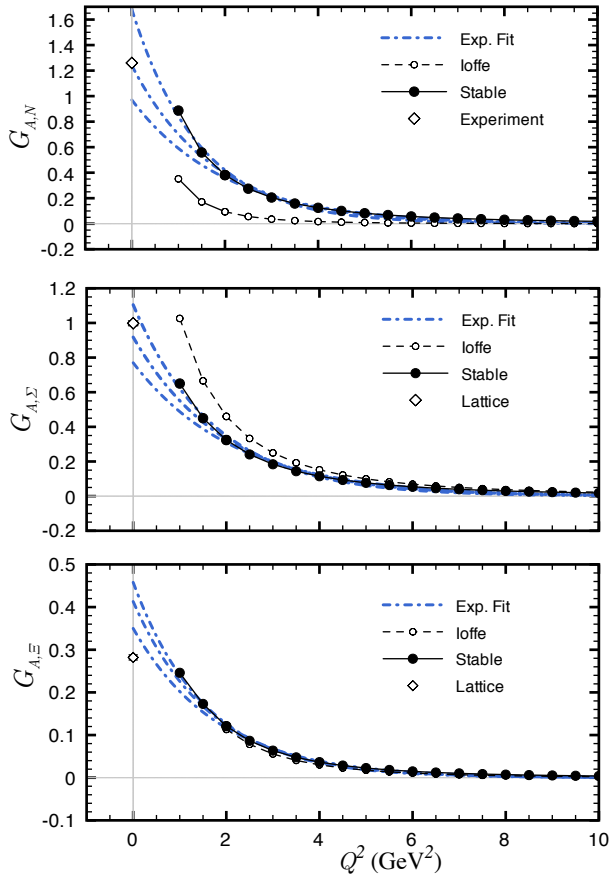}
	\caption{\label{gASX} $G_{A,B}(Q^2)$ of $N$, $\Sigma$ and $\Xi$ as a function of $Q^2$ for the Ioffe current (dashed) and for the stable region of mixing parameter (solid). The diamonds mark the lattice-QCD results for $G_{A,B}(0)$, namely axial charges of the $N$, $\Sigma$ and $\Xi$. The dot-dashed curves show the fit function to an exponential form from three regions: $Q^2>1$~GeV$^2$ (upper), $Q^2>1.5$~GeV$^2$ (middle) and $Q^2>2$~GeV$^2$ (lower).}
\end{figure}	

The sum rules include several parameters that need to be determined. The continuum threshold value for the nucleon is pretty much fixed at $s_0\sim$ 2.25~GeV$^2$ in the literature also from a mass analysis. We choose the values $s_0\sim$ 2.5 and 2.7~GeV$^2$, respectively for $\Sigma$ and $\Xi$. In order to see the dependence of the form factors on the continuum threshold, we vary the values of $s_0$ within a 10~$\%$ region, which leads to a change of less than 10~$\%$ in the final results.

\begin{figure}[th]
	\includegraphics[width=0.5\textwidth]{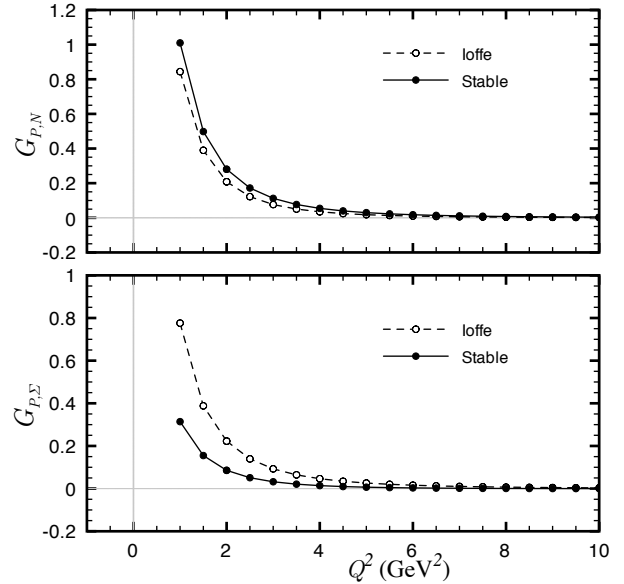}
	\caption{\label{gPSX} Same as Fig.~\ref{gASX} but for $G_{P,B}$ of $N$ and $\Sigma$.}
\end{figure}	

The form factors should be independent of the Borel parameter $M^2$. We consider the regions 1~GeV$^2$ $\leq M^2 \leq$ 2~GeV$^2$ for the nucleon and 2~GeV$^2$ $\leq M^2 \leq$ 4~GeV$^2$ for $\Sigma$ and $\Xi$. We observe that the sum rules are almost independent of $M^2$ in this region; a variation in this region leads to change of the order of 1~$\%$ in the final results. Hence we give our numerical results at $M^2=$2~GeV$^2$ for nucleon and at $M^2=$3~GeV$^2$ for $\Sigma$ and $\Xi$.

The next task is to determine the optimal mixing parameter $\beta$. In the ideal case, the sum rules and hadron properties are independent of this parameter. In order to see if we can achieve such an independence, in Fig.~\ref{costhq} we plot the form factors as a function of $\cos\theta$, where we make a reparameterization using $\beta=\tan\theta$. We explicitly mark the point for Ioffe current, which corresponds to a choice $\beta=-1$. It is observed that a stability region with respect to a change in the mixing parameter can be found around $\cos\theta\sim$0. In further analysis, we concentrate on this stable region and compare the results with those obtained using Ioffe current.

In Fig.~\ref{gASX}, we plot the $G_{A,B}(Q^2)$ of $N$, $\Sigma$ and $\Xi$ as a function of $Q^2$ in the region  $Q^2\geq$1~GeV$^2$~~\footnote{The predictions of LCSR are not reliable at $Q^2 \simeq 0$, but are reliable for $Q^2$ larger then a few GeV$^2$.}, for the Ioffe current ($\beta=-1$) and for the stable region of mixing parameter ($\cos\theta\sim 0$). The qualitative behavior of the form factors agree with our expectations: The values of the axial-vector couplings fall off quickly as we increase the momentum transfer. While there is a considerable discrepancy between the Ioffe and the stable regions for nucleon form factors at low momentum transfers, the results for the form factors are very close to each other in the case of $\Sigma$ and $\Xi$. Particularly for $\Xi$ form factor the two regions produce practically the same results.

For comparison, we also give the lattice-QCD results for $G_{A,B}(0)$, namely axial charges of the $N$, $\Sigma$ and $\Xi$~\cite{Erkol201036}. It was found in Ref.~\cite{Erkol201036} that the axial charges have rather weak quark-mass dependence and the breaking in SU(3)-flavor symmetry is small. Furthermore, the QCDSR results are not yet precise enough to resolve the small variation of  axial charges as a function of quark mass in available lattice-QCD data. Therefore we show the values from SU(3)-flavor symmetric point only. We also note that regarding the signs of the form factors we adopt the convention used in Ref.~\cite{Erkol201036}.

$G_A$ is usually parameterized in terms of a dipole form
\begin{equation}\label{dipform}
	G_{A,B}(Q^2)=g_{A,B}/(1+Q^2/\Lambda_B^2)^2.
\end{equation}
A global average of the nucleon axial mass as determined from neutrino scattering by Budd {\it et al.}~\cite{Budd:2003wb}, $\Lambda_N=1.001\pm 0.020$~GeV, is in good agreement with the theoretically corrected value from pion electroproduction as $\Lambda_N=1.014\pm 0.016$~GeV~\cite{Bernard:2001rs}. A different prediction is made by the K2K Collaboration from quasielastic $\nu_\mu n\rightarrow \mu^- p$ scattering as $\Lambda_N=1.20\pm 0.12$~GeV~\cite{Gran:2006jn}. To extrapolate the sum-rules results to low-momentum--transfer region, we have first tried a two-parameter fit to the dipole form. However this procedure fails to give a good description of data. Instead we fix $g_{A,N}$ to the experimental value and make one parameter fit from 2~GeV$^2$ region. Inserting the experimental value $g_{A,N}=1.2694(28)$ for nucleon and fitting to the dipole form in Eq.~\eqref{dipform}, our sum rules in the stable region of $\beta$ produce $\Lambda_N=1.41$~GeV, a value larger than the experimental result. We make a similar analysis for $\Sigma$ and $\Xi$ axial-vector form factors using the lattice-QCD values for $g_{A,\Sigma}$ and $g_{A,\Xi}$ in the dipole form and find $\Lambda_\Sigma=1.49$~GeV and $\Lambda_\Xi=1.56$~GeV. Our results show that axial masses of $\Sigma$ and $\Xi$ are slightly larger than that of nucleon. Note that, in the VMD model, the pole of the form factors is given by the mass of the (axial) vector meson that couples to the current. The lightest axial vector meson has a mass of $m_A=1.23$~GeV~\cite{Nakamura:2010zzi}; hence our results also are larger from the predictions of the VMD model. 

We have also tried to fit to an exponential form, {\it viz.},
\begin{equation}\label{expform}
	G_{A,B}(Q^2)=g_{A,B} \exp[-Q^2/m_{A,B}^2],
\end{equation}
which allows a plausible description of data with a two-parameter fit. In this case we have tried three fit regions, namely, $Q^2>1$~GeV$^2$, $Q^2>1.5$~GeV$^2$ and $Q^2>2$~GeV$^2$. Our results are shown in Fig.~\ref{gASX} and summarized in Table~\ref{fit_table}. The fits from around $Q^2>1.5$~GeV$^2$ region produce the empirical values of $g_{A,B}$ quite successfully in the case of $N$ and $\Sigma$, while we obtain somewhat higher values of $g_{A,\Xi}$ than that from lattice QCD for all fit regions. We also observe that the axial masses are very close to each other, which indicates a possibly small SU(3)-flavor symmetry breaking in consistency with lattice-QCD findings~\cite{Erkol201036}. It will be interesting to compare our sum-rules results to those from lattice QCD with more realistic setups when available in the near future. 

\begin{table}[t]
	\caption{The values of exponential fit parameters, namely $g_{A,B}$ and $m_{A,B}$, of axial form factors. We give the results of fits from three regions. $g_{A,B}$ values are to be compared with the experimental value  $g_{A,N}=1.2694(28)$~\cite{Nakamura:2010zzi} for nucleon and the lattice-QCD results $g_{A,\Sigma}=0.998(14)$ and $g_{A,\Xi}=0.282(6)$~\cite{Erkol201036} in the case of $\Sigma$ and $\Xi$ respectively.}
	\addtolength{\tabcolsep}{2pt}
\begin{tabular}{ccccccc}
		\hline\hline 
		Baryon & Fit Region~(GeV$^2$) & $g_{A,B}$ & $m_{A,B}$~(GeV)& \\[0.5ex]
		\hline 
		 & [1.0-10] & 1.68 & 1.20  &  \\
		$N$ & [1.5-10]& 1.24 & 1.33  &  \\
		  & [2.0-10] & 0.97 & 1.42  &  \\[1ex]
		& [1.0-10] & 1.11 & 1.32  &  \\
		$\Sigma$ & [1.5-10]& 0.92 & 1.40  &  \\
		  & [2.0-10] & 0.77 & 1.48  &  \\[1ex]
		& [1.0-10] & 0.46 & 1.25  &  \\
		$\Xi$ & [1.5-10]& 0.41 & 1.29  &  \\
		  & [2.0-10] & 0.35 & 1.35  &  \\
		\hline\hline
	\end{tabular}
	\label{fit_table}
\end{table}

In Fig.~\ref{gPSX}, we give similar plots for $G_{P,B}(Q^2)$ of $N$ and $\Sigma$ as a function of $Q^2$. The value of $G_{P,\Xi}$ is negligibly small as compared to other form factors (consistent with zero as can also be seen in Fig.~\ref{costhq}) therefore its figure is not shown. The results from Ioffe and the stable regions are very close to each other in the case of $G_{P,\Xi}$, while we observe some discrepancy for $G_{P,\Sigma}$ form factors. $G_{P,B}$ has a stronger $Q^2$ dependence as compared to $G_{A,B}$. Actually, $G_{P,B}$ has a pole around the pion mass and this can explain the difference in the behaviors of two form factors. We have, unfortunately, not been able to obtain a good fit of $G_{P,B}$ to either dipole or exponential functions. This is probably due to rapid increase of $G_{P,B}$ below $Q^2=$1~GeV$^2$, where we do not have reliable sum-rules data.

\section{Conclusions and Outlook}

We have extracted the isovector axial-vector and induced pseudoscalar form factors of octet baryons by employing the LCSR method. These form factors provide information about the shape and the size of the baryons. The values of the hyperon DA's were not known precisely and this prevented the studies on hyperon structure and form factors from QCD for a long time. However, the DA's have been recently calculated up to twist six~\cite{Liu:2008yg, Liu:2009uc}, which allows us to give a description of form factors at high-momentum transfers. Unfortunately, there is no sufficient experimental data yet to compare our results with in this region. However, the new generation higher-energy neutrino experiments, such as Miner$\nu$a~\cite{Drakoulakos:2004gn} will span a wide region of momentum transfers and will probe baryon axial-form factors with high precision in the near future. 

In the low-energy region, we have compared our results with those from experiment and two-flavor lattice QCD simulations~\cite{Erkol201036}. We have observed that there is a nice qualitative and quantitative agreement, which can be suitably reproduced by an exponential form. With the availability of the lattice-QCD data in the low-$Q^2$, as well as in the high-$Q^2$ region, we will be able to give a more accurate comparison of these two complementary approaches. Work along this direction is still in progress, where it is aimed to extract baryon form factors in a wide range of momentum transfers with larger 2+1-flavor lattices of smaller lattice spacing and quark masses. We also aim to extract isoscalar form factors and extend our study to non-diagonal baryon transitions as well. Our work along this direction is also in progress.
\acknowledgments
We gratefully acknowledge very useful discussions with M. Oka and T. T. Takahashi. This work has been supported by The Scientific and Technological Research Council of Turkey (T\"{U}B{\.I}TAK) under project number 110T245. The work of A. O. is also partially supported by the European Union (HadronPhysics2 project Study of strongly interacting matter).

\end{document}